# Predicting protein contact map using evolutionary and physical constraints by integer programming (extended version)

Zhiyong Wang[1] and Jinbo Xu[1,*]

[1]Toyota Technological Institute at Chicago

6045 S Kenwood, IL, 60637, USA

[*]Corresponding author (email: jinboxu@gmail.com)

**ABSTRACT**

**Motivation.** Protein contact map describes the pairwise spatial and functional relationship of residues in a protein and contains key information for protein 3D structure prediction. Although studied extensively, it remains very challenging to predict contact map using only sequence information. Most existing methods predict the contact map matrix element-by-element, ignoring correlation among contacts and physical feasibility of the whole contact map. A couple of recent methods predict contact map based upon residue co-evolution, taking into consideration contact correlation and enforcing a sparsity restraint, but these methods require a very large number of sequence homologs for the protein under consideration and the resultant contact map may be still physically unfavorable.

**Results.** This paper presents a novel method PhyCMAP for contact map prediction, integrating both evolutionary and physical restraints by machine learning and integer linear programming (ILP). The evolutionary restraints include sequence profile, residue co-evolution and context-specific statistical potential. The physical restraints specify more concrete relationship among contacts than the sparsity restraint. As such, our method greatly reduces the solution space of the contact map matrix and thus, significantly improves prediction accuracy. Experimental results confirm that PhyCMAP outperforms currently popular methods no matter how many sequence homologs are available for the protein under consideration. PhyCMAP can predict contacts within minutes after PSIBLAST search for sequence homologs is done, much faster than the two recent methods PSICOV and EvFold.

**Availability**: see http://raptorx.uchicago.edu for the web server.

## 1 INTRODUCTION

In this paper we say two residues of a protein are in contact if their Euclidean distance is less than 8Å. The distance of two residues can be calculated on $C_\alpha$ or $C_\beta$ atoms, i.e., $C_\alpha$ or $C_\beta$ contacts. A protein contact map is a binary $L \times L$ matrix where $L$ is the protein length. In this matrix, an element with value 1 indicates the corresponding two residues are in contact; otherwise, not in contact. Protein contact map describes the pairwise spatial and functional relationship of residues in a protein. Predicting contact map from sequence has been an active research topic in recent years partially because contact map is helpful for protein 3D structure prediction (Ortiz, et al., 1999; Vassura, et al., 2008; Vendruscolo, et al., 1997; Wu, et al., 2011) and protein model quality assessment (Zhou and Skolnick, 2007). Protein contact map has also been used for protein structure alignment and classification (Caprara, et al., 2004; Wang, et al., 2013; Xu, et al., 2006).

Many machine learning methods have been developed for protein contact prediction in the past decades (Fariselli and Casadio, 1999; Göbel, et al., 2004; Olmea and Valencia, 1997; Punta and Rost, 2005; Vendruscolo and Domany, 1998; Vullo, et al., 2006). For example, SVMSEQ (Wu and Zhang, 2008) and SVMcon (Cheng and Baldi, 2007) use support vector machines; NNcon (Tegge, et al., 2009) uses a recursive neural network; Distill (Baú, et al., 2006) uses a 2-dimensional recursive neural network; and CMAPpro (Di Lena, et al., 2012) uses a multi-layer neural network. Although very different, these methods are common in that they predict the contact map matrix element-by-element, ignoring the correlation among contacts and also physical feasibility of the whole contact map (physical constraints are not totally independent of contact correlation). A special type of physical constraint is that a contact map matrix must be sparse, i.e., the number of contacts in a protein is only linear in its length.

Two recent methods (PSICOV (Jones, et al., 2012) and Evfold (Morcos, et al., 2011)) predict contacts using only residue co-evolution information derived from sequence homologs and enforcing the above-mentioned sparsity constraint. However, both of them require a large number (at least several hundreds) of sequence homologs for the protein under prediction. This makes the predicted contacts not very useful in real-world protein modeling since a (globular) protein with many sequence homologs usually has very similar templates in PDB and thus, template-based models are of good quality and cannot be further improved by predicted contacts. Conversely, a protein without close templates in PDB, which may benefit from contact prediction, usually has very few sequence homologs. Further, these two methods enforce only a very simple sparsity constraint (i.e., the total number of contacts in a protein is small), ignoring many more concrete constraints. To name a few, one residue can have only a small number of contacts, depending on its secondary structure and neighboring residues. The number of contacts between two beta strands is bounded by the strand length.

Astro-Fold (Klepeis and Floudas, 2003) possibly is the first contact prediction method that makes use of physical constraints, which imply the sparsity constraint used by PSICOV and Evfold. However, some of the physical constraints are too restrictive and possibly unrealistic. For example, it requires that a residue in one beta strand can only be in contact with one residue in another beta strand. More importantly, Astro-Fold does not make use of evolutionary information and thus, significantly reduces its prediction accuracy.

In this paper, we present a novel method PhyCMAP for contact map prediction by integrating both evolutionary and physical constraints using machine learning (i.e., Random Forests) and integer linear programming (ILP). PhyCMAP first predicts the probability of any two residues forming a contact using both evolutionary and non-evolutionary information. Then, PhyCMAP infers a given number of top contacts based upon predicted contact probabilities by enforcing a set of realistic physical constraints, which specify more concrete relationship among contacts and also imply the sparsity restraint used by PSICOV and Evfold. By combining both evolutionary and physical constraints, our method greatly reduces the solution space of a contact map and leads to much better prediction accuracy. Experimental results confirm that PhyCMAP outperforms currently popular methods no matter how many sequence homologs are available for the protein under prediction.

## 2 METHODS

**Overview**. As shown in Figure 1, our method consists of several key components. First, we employ Random Forests (RF) to predict the contact probability of any two residues from both evolutionary and non-evolutionary information. Then we employ an integer linear programming (ILP) method to select a set of top contacts by maximizing their accumulative probabilities subject to a set of physical constraints, which are represented as linear constraints. The resultant top contacts form a physically favorable contact map for the protein under consideration.

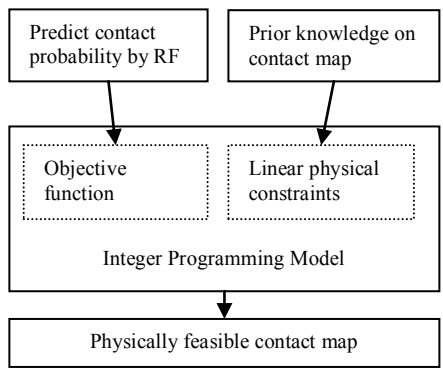

**Figure 1.** The overview of our contact prediction approach.

### 2.1 Predicting contact probability by Random Forests

We use Random Forests (RF) to predict the probability of any two residues forming a contact from several types of information: 1) evolutionary information from a single protein family, e.g., PSI-BLAST sequence profile (Altschul and Koonin, 1998) and two types of mutual information; 2) non-evolutionary information derived from the solved protein structures, e.g., residue contact potential; 3) mixed information from the above two sources, e.g., EPAD (a context-specific distance-based statistical potential) (Zhao and Xu, 2012), secondary structure predicted by PSIPRED (Jones, 1999), and homologous pairwise contact score; and 4) amino acid physicochemical properties. Some features are calculated on the residues in a local window of size 5 centered at the residues under consideration. In total there are ~300 features for each residue pair. We trained our RF model using the RandomForest package in R (Breiman, 2001; Liaw and Wiener, 2002) and selected the model parameters by 5-fold cross validation. Below we briefly explain some important features.

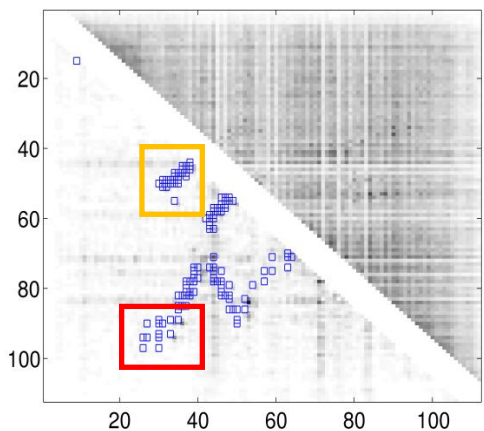

**Figure 2.** The contrastive mutual information (lower triangle) and mutual information (upper triangle) of protein 1j8bA.The native contacts are represented by boxes. The big yellow and red boxes contain a group of medium-range and long-range contacts, respectively.

**Sequence profile.** The position-specific mutation scores at residues $i$ and $j$ and their neighboring residues are used.

**Contrastive mutual information (CMI).** Let $m_{i,j}$ denote the mutual information (MI) between two residues $i$ and $j$, which can be calculated from the multiple sequence alignment (MSA) of all the sequence homologs. We define the contrastive mutual information (CMI) as the MI difference between one residue pair and all of its neighboring pairs, which can be calculated as follows.

$$CMI_{i,j} = (m_{i,j} - m_{i-1,j})^2 + (m_{i,j} - m_{i+1,j})^2 + (m_{i,j} - m_{i,j-1})^2 + (m_{i,j} - m_{i,j+1})^2$$

The CMI measures how the co-mutation strength of one residue pair differs from its neighboring pairs. By using CMI instead of MI, we can remove the background bias of MI in a local region, as shown in Figure 2. When only a small number of sequence homologs are available (or at some conserved positions with entropy <0.3), we may have very low MI, which may result in artificially high CMI. To avoid this, we directly set the CMI in these situations to 0.

**Mutual information for the chaining effect of residue coupling.** When residue A has strong interaction with B and B has strong interaction with residue C, it is likely that residue A also has interaction with C. To account for this kind of chaining effect of residue coupling, we use the powers of the MI matrix. In particular, we use $MI^k$ where $k$ ranges from 2 to 11 to capture this kind of chaining effect. When there are many sequence homologs (i.e., MI is accurate), this kind of MI powers greatly improves medium- and long-range contact prediction (see Results for details).

**Residue contact potential.** We use residue contact-based potential described in (Tan, et al., 2006)

**EPAD.** It is a context-specific distance-dependent statistical potential (Zhao and Xu, 2012), derived from protein evolutionary information and solved protein structures in PDB. The $C_\alpha$ and $C_\beta$ atomic interaction potential at all the distance bins is used as features. The atomic distance is discretized into bins by 1Å and all the distance>15Å is grouped into a single bin.

**Homologous pairwise contact score (HPS).** This score quantify the probability of a residue pair forming a contact between two secondary structures. Let $PS_{beta}$ *(a, b)* denote the probability of two amino acids *a* and *b* forming a beta contact. Let $PS_{helix}$ *(a, b)* denote the probability of two amino acids *a* and *b* forming a helix contact. Both $PS_{beta}$ *(a, b)* and $PS_{helix}$ *(a, b)* are derived from a simple statistics on PDB25. Let H denote the set of sequence homologs of the protein under consideration. Let *i* and *j* denote two positions in the multiple sequence alignment of the homologs. We calculate the homologous pairwise contact score *HPS* for the two positions *i* and *j* as follows.

$$PS(i,j) = \frac{1}{|H|}(\sum_{h \in H} PS_{beta}\left(h_i, h_j\right) \text{ or } \sum_{h \in H} PS_{helix}(h_i, h_j))$$

Where $h_i$ and $h_j$ denote the amino acids of the homolog *h* at positions *i* and *j*, respectively.

## 2.2 The integer linear programming (ILP) method

**The variables.** Let *i* and *j* denote residue positions and L the protein length. Let *u* and *v* index secondary structure segments of a protein. Let *Begin(u)* and *End(u)* denote the first and last residues of the segment *u* and *SSeg(u)* the set $_{\{i \mid Begin(u) \leq i \leq End(u)\}}$. Let *SStype(u)* denote the secondary structure type of one residue or one segment *u*. Let *Len(u)* denote the length of the segment *u*. We use the binary variables in Table 1.

**Table 1.** The binary variables used in the ILP formulation.

| Variable | Explanation |
| --- | --- |
| $X_{i,j}$ | equal to 1 if there is a contact between two residues *i* and *j*. |
| $AP_{u,v}$ | equal to 1 if two beta-strands *u* and *v* form an anti-parallel beta-sheet. |
| $P_{u,v}$ | equal to 1 if two beta-strands *u* and *v* form a parallel beta-sheet. |
| $S_{u,v}$ | equal to 1 if two beta-strands *u* and *v* form a beta-sheet. |
| $T_{u,v}$ | equal to 1 if there is an alpha-bridge between two helices *u* and *v*. |
| $R_r$ | a non-negative integral relaxation variable of the $r^{th}$ soft constraint. |

**The objective function.** Intuitively we shall choose those contacts with the highest probability predicted by our Random Forest model, i.e., the objective function shall be the sum of predicted probabilities of the selected contacts. However, the selected contacts shall also minimize the violation of the physical constraints. To enforce this, we use a vector of relaxation variables *R* to measure the degree to which all the soft constraints are violated. Thus, our objective function is as follows.

$$\max_{X,R} \sum_{j-i \geq 6} X_{i,j} \times P_{i,j} - g(R)$$

Where $P_{i,j}$ is the contact probability predicted by our Random Forest model for two residues and $g(R) = \sum_r R_r$ is a linear penalty function where *r* run over all the soft constraints. The relaxation variables will be further explained in the following section.

**The constraints.** We use both soft and hard constraints. There is a single relaxation variable for each group of soft constraint, but the hard constraints are strictly enforced. We penalize the violation of soft constraints by incorporating the relaxation variables to the objective function. The constraints in groups 1, 2, and 6 are soft constraints. Those in groups 3, 4, 5 and 7 are hard constraints, some of which are similar to what are used by Astro-Fold (Klepeis and Floudas, 2003).

*Group 1.* This group of soft constraints bound from above the total number of contacts that can be formed by a single residue *i* (in a secondary structure type s1) with all the other residues in a secondary structure type s2.

$$\forall i, SStype\ (i) = s1, \sum_{j:SStype\ (j)=s2} X_{i,j} \leq R_1 + b_{s1,s2}$$

Where $b_{s1,s2}$ is a constant empirically determined from our training data (see Table 2) and $R_1$ is the relaxation variable.

**Table 2.** The empirical values of $b_{s1,s2}$ calculated from the training data. The first column indicates the combination of two secondary structure types: H(α-helix), E(β-strand) or C(coil). Each row contains two empirical values for a pair of secondary structure types. Column "95%": 95% of the secondary structure pairs have the number of contacts smaller than the value in this column; Column "Max": the largest number of contacts.

| s1,s2 | 95% | Max |
|-------|-----|-----|
| H,H   | 5   | 12  |
| H,E   | 3   | 10  |
| H,C   | 4   | 11  |
| E,H   | 4   | 12  |
| E,E   | 9   | 13  |
| E,C   | 6   | 15  |
| C,H   | 3   | 12  |
| C,E   | 5   | 12  |
| C,C   | 6   | 20  |

*Group 2.* This group of constraints bound the total number of contacts between two strands sharing at least one contact. Let $u$ and $v$ denote two beta strands. We have

$$\sum_{i \in SSeg \ (v), \ j \in SSeg \ (u)} X_{i,j} + R_2 \geq 3 \times S_{u,v} \times \min(\ Len \ (u), \ Len \ (v))$$

$$\sum_{i \in SSeg \ (v), \ j \in SSeg \ (u)} X_{i,j} \geq 3.3 \times \max(\ Len \ (u), \ Len \ (v)) + R_3$$

The two constraints are explained in Figure 3 as follows. Figure 3(A) shows that the total number of contacts between two beta strands diverges into two groups when $\min(Len(u), Len(v)) \geq 9$. One group is due to beta strand pairs forming a beta sheet. The other may be due to incorrectly predicted beta strands or beta strand pairs not in a beta sheet. Figure 3(B) shows that the total number of contacts between a pair of beta strands has an upper bound proportional to the length of the longer beta strand. Since there are points below the skew line in Figure 3(A), which indicate that a beta strand pair may have fewer than $3 \times \min(Len(u), Len(v))$ contacts, we add a relaxation variable $R_2$ to the lower bound constraints in Group 2. Similarly, we use a relaxation variable $R_3$ for the upper bound constraints.

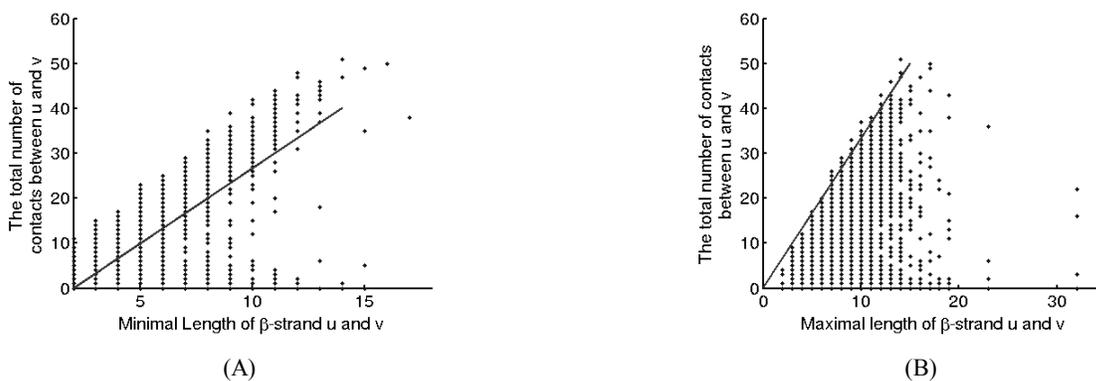

(A)                                                    (B)

**Figure 3.** The skew lines indicate the bounds for the total number of contacts between two beta strands. (A) lower bound; (B) upper bound.

*Group 3.* When two strands form an anti-parallel beta-sheet, the contacts of neighboring residue pairs shall satisfy the following constraints.

$$X_{i,j} \geq X_{i-1,j+1} + X_{i+1,j-1} - 1$$

Where $i, i \pm 1$ are residues in one strand and $j, j \pm 1$ are residues in the other strand.

*Group 4.* When there are parallel contacts between two strands, the contacts of neighboring residue pairs should satisfy the following constraints.

$$X_{i,j} \geq X_{i-1,j-1} + X_{i+1,j+1} - 1$$

Where $i, i \pm 1$ are residues in one strand and $j, j \pm 1$ are residues in the other strand.

*Group 5.* One beta-strand $u$ can form beta-sheets with up to 2 other beta-strands.

$$\sum_{v:SStype\ (v)=beta} S_{u,v} \leq 2$$

*Group 6.* There is no contact between the start and end residues of a loop segment $u$.

$$X_{i,j} \leq 0 + R_4, i = Begin\ (u), j = End\ (u)$$

In our training set, there are totally ~8000 loop segments, and only 3.4% of them have a contact between the start and end residues. To allow the rare cases, we use a relaxation variable $R_4$ in the constraints.

*Group 7.* One residue $i$ cannot have contacts with both $j$ and $j+2$ when $j$ and $j+2$ are in the same alpha helix.

$$X_{i,j} + X_{i,j+2} \leq 1$$

*Group 8.* This group of constraints models the relationship among different groups of variables.

$$AP_{u,v} + P_{u,v} = S_{u,v}$$

$$X_{i,j} \leq S_{u,v}, \forall i \in SSeg\ (u), j \in SSeg\ (v)$$

$$\sum_{1 \leq i < j \leq L, j-i \geq 6} X_{i,j} = k$$

Where $k$ is the number of top contacts we want to predict.

**Integer program solver:** IBM CPLEX academic version V12.1 (CPLEX, 2009).

**Training data.** It consists of 900 non-redundant protein structures, any two of which share no more than 25% sequence identity. Since there are far fewer contacts than non-contacts, we use all the contacts and randomly sample only 20% of the non-contacts as the training data. All the training proteins are selected before CASP10 (the 10[th] Critical Assessment of Structure Prediction) started in May 2012.

**Test data I: CASP10.** This set contains 123 CASP10 targets with publicly available native structures. Meanwhile, 36 of them are classified as hard targets because the top half of their submitted models have average TMscore<0.5. When they were just released, most of the CASP10 targets share low sequence identity (<25%) with the proteins in PDB. BLAST indicates that there are only 5 short CASP10 targets (~50 residues) which have sequence identity slightly above 30% with our training proteins.

**Test data II: Set600.** This set contains 601 proteins randomly extracted from PDB25 (Brenner, et al., 2000) and was constructed before CASP10 started. They have the following properties: 1) less than 25% sequence identity with the training proteins; 2) each protein has at least 50 residues and an X-ray structure with resolution better than 1.9Å; 3) each protein has at least 5 residues with predicted secondary structure being alpha-helix or beta-strand.

Both the training set and Set600 are sampled from PDB25 (Wang and Dunbrack, 2003), in which any two proteins share less than 25% sequence identity. Sequence identity is calculated using the method in (Brenner, et al., 2000).

**Programs to be compared.** We compare our method, denoted as PhyCMAP, with four state-of-the-art methods: CMAPpro (Di Lena, et al., 2012), NNcon (Tegge, et al., 2009), PSICOV (Jones, et al., 2012) and Evfold (Morcos, et al., 2011). We run NNCon, PSICOV and Evfold locally and CMAPpro by its web server. We do not compare our method with Astro-Fold because it is not publicly available. Since Astro-Fold does not make use of any evolutionary information, it does not perform well for many proteins.

**Evaluation methods.** Depending on the chain distance of the two residues, there are three kinds of contacts: short-range, medium-range and long-range. Short-range contacts are closely related to local conformation and are relatively easy to predict. Medium-range and long-range contacts determine the overall shape of a protein and are more challenging to predict. We evaluate prediction accuracy using the top 5, L/10, L/5 predicted contacts where L is the protein length.

**$M_{eff}$: the number of non-redundant sequence homologs.** Given a target protein and the multiple sequence alignment of all of its homologs, we measure the number of non-redundant sequence homologs by $M_{eff}$ as follows.

$$M_{eff} = \sum_i \frac{1}{\sum_j S_{i,j}} \qquad (1)$$

Where both $i$ and $j$ go over all the sequence homologs and $S_{i,j}$ is a binary similarity value between two proteins. Following Evfold (Morcos, et al., 2011), we compute the similarity of two sequence homologs using their hamming distance. That is, $S_{i,j}$ is 1 if the normalized hamming distance is < 0.3; 0 otherwise.

## 3 RESULTS

**Performance on the CASP10 set.** As shown in Table 3, tested on the whole CASP10 set, our PhyCMAP significantly exceeds the 2[nd] best method CMAPpro in terms of the accuracy of the top 5, L/10 and L/5 predicted contacts. The advantage of PhyCMAP over CMAPpro becomes smaller but still substantial when short-range contacts are excluded from consideration. PhyCMAP significantly outperforms NNcon, PSICOV and Evfold no matter how the performance is evaluated. Figure 4 shows the head-to-head comparison between PhyCMAP and CMAPpro and PSICOV for long-range contact prediction.

**Table 3.** This table lists the prediction accuracy of PhyCMAP, PSICOV, NNCON, CMAPpro and Evfold for short-, medium- and long-range contacts, tested on 123 CASP10 targets.

| | Short-range, sequence distance from 6 to 12 | | | Medium and long-range, sequence distance > 12 | | | Medium range, sequence distance > 12 and ≤24 | | | Long range, sequence distance > 24 | | |
|---|---|---|---|---|---|---|---|---|---|---|---|---|
| | Top 5 | L/10 | L/5 | Top 5 | L/10 | L/5 | Top 5 | L/10 | L/5 | Top 5 | L/10 | L/5 |
| PhyCMAP($C_\alpha$) | 0.623 | 0.555 | 0.459 | 0.650 | 0.584 | 0.523 | 0.585 | 0.518 | 0.448 | 0.421 | 0.363 | 0.320 |
| PhyCMAP($C_\beta$) | 0.667 | 0.580 | 0.482 | 0.655 | 0.604 | 0.539 | 0.621 | 0.550 | 0.465 | 0.514 | 0.425 | 0.373 |
| PSICOV($C_\alpha$) | 0.294 | 0.225 | 0.179 | 0.397 | 0.345 | 0.306 | 0.384 | 0.303 | 0.255 | 0.350 | 0.277 | 0.226 |
| PSICOV($C_\beta$) | 0.379 | 0.281 | 0.223 | 0.522 | 0.458 | 0.405 | 0.515 | 0.387 | 0.316 | 0.457 | 0.372 | 0.315 |
| NNCON($C_\alpha$) | 0.595 | 0.499 | 0.399 | 0.472 | 0.409 | 0.358 | 0.463 | 0.393 | 0.334 | 0.286 | 0.239 | 0.188 |
| CMAPpro($C_\alpha$) | 0.506 | 0.437 | 0.368 | 0.517 | 0.466 | 0.424 | 0.485 | 0.414 | 0.363 | 0.380 | 0.336 | 0.297 |
| CMAPpro($C_\beta$) | 0.543 | 0.477 | 0.395 | 0.519 | 0.466 | 0.415 | 0.491 | 0.419 | 0.370 | 0.395 | 0.347 | 0.313 |
| Evfold($C_\alpha$) | 0.236 | 0.193 | 0.165 | 0.380 | 0.326 | 0.295 | 0.351 | 0.294 | 0.249 | 0.328 | 0.257 | 0.225 |

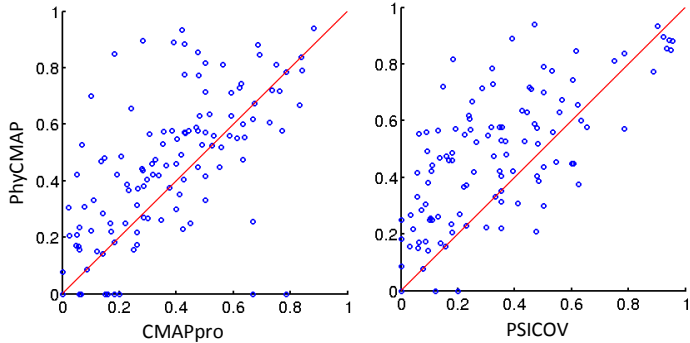

Figure 4. Head-to-head comparison of PhyCMAP, CMAPpro and PSICOV on the 123 CASP10 targets. Each point represents the accuracy of two methods under comparison on a single test protein. Top L/5 predicted long-range $C_\beta$ contacts are considered. The average accuracies of PhyCMAP, CMAPpro, and PSICOV are 0.373, 0.313, and 0.315, respectively.

**Performance on the 36 hard CASP10 targets.** As shown in Table 4, tested on the 36 hard CASP10 targets our PhyCMAP exceeds the 2[nd] best method CMAPpro by 5-7% in terms of the accuracy of the top 5, L/10 and L/5 predicted contacts. The advantage of PhyCMAP over CMAPpro becomes smaller but still substantial when short-range contacts are excluded from consideration. PhyCMAP significantly outperforms NNcon, PSICOV and Evfold no matter how many predicted contacts are evaluated. PSICOV and Evfold almost fail on these hard CASP10 targets. By contrast, CMAPpro, NNcon and PhyCMAP still can correctly predict some contacts. Figure 5 shows the head-to-head comparison of PhyCMAP, CMAPpro and PSICOV for medium- and long-range contact prediction.

**Table 4.** Accuracy of PhyCMAP, PSICOV, NNcon, Evfold and CMAPpro on the 36 hard CASP10 targets. The $C_\beta$ results are in gray rows.

| | Short-range, sequence distance from 6 to 12 | | | Medium and long-range, sequence distance > 12 | | |
|---|---|---|---|---|---|---|
| | Top 5 | L/10 | L/5 | Top 5 | L/10 | L/5 |
| PhyCMAP ($C_\alpha$) | 0.456 | 0.439 | 0.378 | 0.394 | 0.378 | 0.325 |
| PhyCMAP ($C_\beta$) | 0.478 | 0.469 | 0.414 | 0.444 | 0.409 | 0.363 |
| PSICOV ($C_\alpha$) | 0.100 | 0.083 | 0.082 | 0.183 | 0.156 | 0.150 |
| PSICOV ($C_\beta$) | 0.144 | 0.113 | 0.103 | 0.239 | 0.196 | 0.180 |

| | | | | | |
|---|---|---|---|---|---|
| NNCON ($C_\alpha$) | 0.400 | 0.372 | 0.320 | 0.367 | 0.317 | 0.289 |
| CMAPpro ($C_\alpha$) | 0.383 | 0.347 | 0.314 | 0.328 | 0.322 | 0.292 |
| CMAPpro ($C_\beta$) | 0.433 | 0.398 | 0.344 | 0.394 | 0.362 | 0.308 |
| Evfold ($C_\alpha$) | 0.100 | 0.095 | 0.094 | 0.194 | 0.179 | 0.156 |

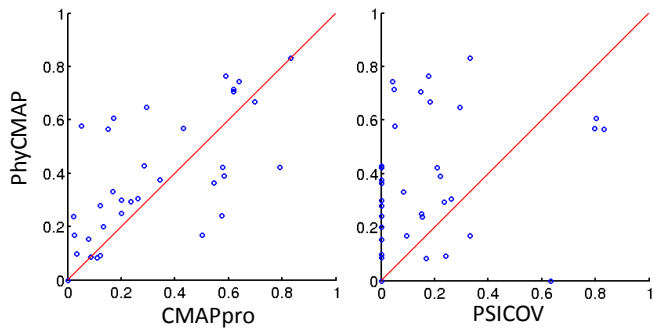

**Figure 5.** Head-to-head comparison of PhyCMAP, CMAPpro and PSICOV on the 36 hard CASP10 targets. Each point represents the accuracy of two methods under comparison on a single test protein. Top L/5 predicted medium- and long-range $C_\beta$ contacts are considered. The average accuracies of PhyCMAP, CMAPpro, and PSICOV are 0.363, 0.308, and 0.180, respectively.

Note that both PSICOV and Evfold, two recent methods receiving a lot of attentions from the community, do not perform very well on the CASP10 set. This is partially because they require a large number of sequence homologs for the protein under prediction. Nevertheless, most of the CASP targets, especially the hard ones, do not satisfy this requirement because a protein with so many homologs very likely has highly similar templates in PDB and thus, was not used as a CASP target.

**Prediction accuracy vs. the number of sequence homologs.** We divide the 123 CASP10 targets into 5 groups according to their $\log M_{eff}$ values: [0,2), [2,4), [4,6), [6,8), [8,10], which contain 19, 17, 25, 36 and 26 targets, respectively. Meanwhile $M_{eff}$ is the number of non-redundant sequence homologs for the protein under consideration (see Methods for definition). Only medium- and long-range contacts are considered here. Figure 6 clearly shows that the prediction accuracy increases with respect to $M_{eff}$. The more non-redundant homologs are available, the better prediction accuracy can be achieved by PhyCMAP, Evfold and PSICOV. However, CMAPpro and NNCON have decreased accuracy when $\log M_{eff} > 8$. Figure 6 also shows that PhyCMAP outperforms Evfold, CMAPpro and NNCON across all $M_{eff}$. PhyCMAP greatly outperforms PSICOV in predicting $C_\alpha$ contacts regardless of $M_{eff}$ and also in predicting $C_\beta$ contacts when $\log M_{eff} \leq 6$. PhyCMAP is comparable to PSICOV in predicting $C_\beta$ contacts when $\log M_{eff} > 6$.

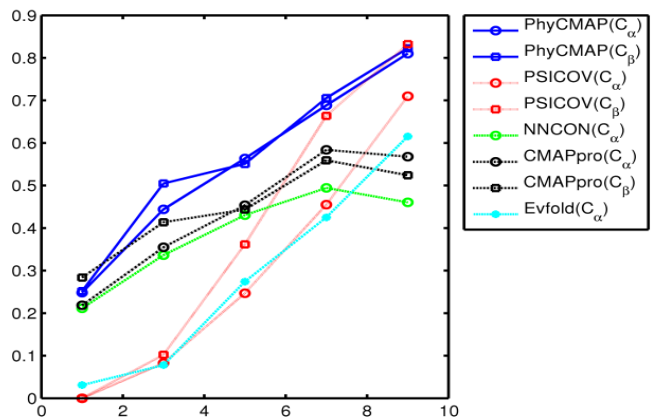

Figure 6. The relationship between prediction accuracy and the number of non-redundant sequence homologs ($M_{eff}$). X-axis is $\log M_{eff}$ and Y-axis is the mean accuracy of top L/10 predicted contacts in the corresponding CASP10 target group. Only medium- and long-range contacts are considered.

**Performance on Set600.** To fairly compare our method with Evfold (Morcos, et al., 2011) and PSICOV (Jones, et al., 2012), both of which require a large number of sequence homologs, we divide Set600 into two subsets based upon $M_{eff}$, the number of non-redundant sequence homologs. The first subset contains 471 proteins with $M_{eff} > 100$ (see Methods for definition). Each protein in this subset has >500 sequence homologs, which satisfies the requirement of PSICOV. The 2nd subset is more challenging, containing 130 proteins with $M_{eff} \leq 100$. As shown in Table 5, even on the 1st subset, PhyCMAP still exceeds PSICOV and Evfold, although the advantage over PSICOV is not substantial for $C_\beta$ contact prediction when short-range contacts are excluded from consideration. PhyCMAP also outperforms NNcon and CMAPpro on this set. As shown in Table 6, on the 2nd subset PhyCMAP significantly outperforms PSICOV and is slightly better than CMAPpro and NNcon. Figures 7 and 8 show the head-to-head comparison of PhyCMAP, PSICOV and CMAPpro. As shown in the figures, when $M_{eff} > 100$, PhyCMAP outperforms CMAPpro on most test proteins; when $M_{eff} \leq 100$, PhyCMAP outperforms both CMAPpro and PSICOV on most test proteins. These results again confirm that our method applies to a protein without many sequence homologs, on which PSICOV and Evfold usually fail.

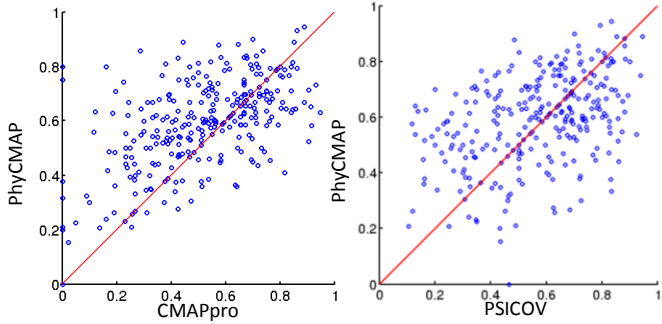

Figure 7. Head-to-head comparison of PhyCMAP, CMAPpro and PSICOV on the Set600 proteins with $M_{eff}$>100. Each point represents the accuracy of two methods under comparison on a single test protein. Top L/5 predicted medium- and long-range $C_\beta$ contacts are considered. The average accuracies of Phy-CMAP, CMAPpro, and PSICOV are 0.611, 0.515, and 0.569, respectively

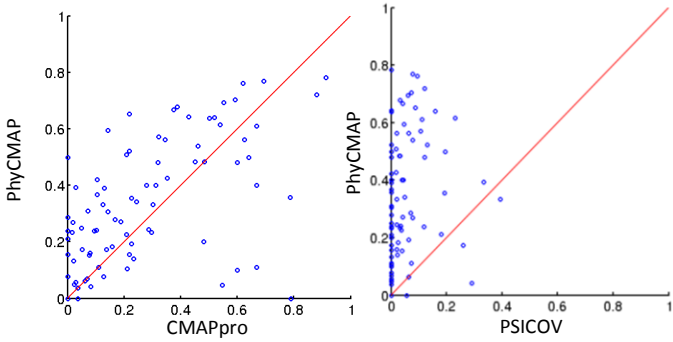

Figure 8. Head-to-head comparison of PhyCMAP, CMAPpro and PSICOV on the Set600 proteins with $M_{eff}$≤100. Each point represents the accuracy of two methods under comparison on a single test protein. Top L/5 predicted medium- and long-range $C_\beta$ contacts are considered. The average accuracies of Phy-CMAP, CMAPpro, and PSICOV are 0.345, 0.287, and 0.059, respectively.

**Table 5.** Accuracy on the 471 proteins with $M_{eff} > 100$.

|  | Short-range, sequence distance from 6 to 12 | | | Medium and long-range, sequence distance > 12 | | |
|---|---|---|---|---|---|---|
|  | Top 5 | L/10 | L/5 | Top 5 | L/10 | L/5 |
| PhyCMAP ($C_\alpha$) | 0.761 | 0.653 | 0.539 | 0.716 | 0.675 | 0.611 |
| PhyCMAP ($C_\beta$) | 0.746 | 0.637 | 0.531 | 0.731 | 0.656 | 0.608 |
| PSICOV ($C_\alpha$) | 0.457 | 0.341 | 0.257 | 0.528 | 0.465 | 0.411 |
| PSICOV ($C_\beta$) | 0.584 | 0.425 | 0.316 | 0.732 | 0.646 | 0.565 |
| NNCON ($C_\alpha$) | 0.512 | 0.432 | 0.355 | 0.432 | 0.361 | 0.308 |
| CMAPpro ($C_\alpha$) | 0.682 | 0.551 | 0.431 | 0.710 | 0.642 | 0.574 |
| CMAPpro ($C_\beta$) | 0.671 | 0.542 | 0.436 | 0.674 | 0.601 | 0.532 |
| Evfold ($C_\alpha$) | 0.379 | 0.297 | 0.234 | 0.473 | 0.438 | 0.400 |

**Table 6.** Accuracy on the 130 proteins with $M_{eff} \leq 100$.
Evfold is not listed since it does not produce meaningful predictions for some proteins.

|  | Short-range, sequence distance from 6 to 12 | | | Medium and long-range, sequence distance > 12 | | |
|---|---|---|---|---|---|---|
|  | Top 5 | L/10 | L/5 | Top 5 | L/10 | L/5 |
| PhyCMAP ($C_\alpha$) | 0.534 | 0.451 | 0.377 | 0.432 | 0.372 | 0.319 |
| PhyCMAP ($C_\beta$) | 0.505 | 0.435 | 0.365 | 0.418 | 0.364 | 0.314 |
| PSICOV ($C_\alpha$) | 0.060 | 0.061 | 0.064 | 0.049 | 0.039 | 0.035 |
| PSICOV ($C_\beta$) | 0.077 | 0.070 | 0.073 | 0.069 | 0.050 | 0.045 |
| NNCON ($C_\alpha$) | 0.442 | 0.363 | 0.309 | 0.368 | 0.339 | 0.301 |
| CMAPpro ($C_\alpha$) | 0.435 | 0.365 | 0.314 | 0.368 | 0.331 | 0.300 |
| CMAPpro (C$_\beta$) | 0.532 | 0.434 | 0.353 | 0.358 | 0.331 | 0.280 |

Note that CMAPpro used Astral 1.73 (Brenner, et al., 2000; Di Lena, et al., 2012) as its training set, which shares >90% sequence identity with 226 proteins in Set600 (180 with $M_{eff}$>100 and 46 with $M_{eff}$≤100). To more fairly compare the prediction methods, we exclude the 226 proteins from Set600 that share >90% sequence identity with the CMAPpro training set. Here the sequence identity is calculated using CD-HIT (Li and Godzik, 2006; Li, et al., 2001). This results in a set of 291 proteins with $M_{eff}$ > 100 and 84 proteins $M_{eff}$≤100. Table 7 shows that PhyCMAP greatly outperforms CMAPpro and Evfold on the reduced dataset. PhyCMAP also outperforms PSICOV in predicting C$_\alpha$ contacts, but is slightly worse in predicting long-range C$_\beta$ contacts.

**Table 7.** This table lists the prediction accuracy of PhyCMAP, PSICOV, NNCON, CMAPpro and Evfold for short-, medium- and long-range contacts, tested on a reduced Set600.

| | Short-range, sequence distance from 6 to 12 | | | Medium and long-range, sequence distance > 12 | | | Medium range, sequence distance > 12 and ≤24 | | | Long range, sequence distance > 24 | | |
|---|---|---|---|---|---|---|---|---|---|---|---|---|
| Accuracy on the 291 Set600 proteins with $M_{eff}$>100 and ≤90% sequence identify with Astral 1.73 | | | | | | | | | | | | |
| | Top 5 | L/10 | L/5 | Top 5 | L/10 | L/5 | Top 5 | L/10 | L/5 | Top 5 | L/10 | L/5 |
| PhyCMAP(C$_\alpha$) | 0.759 | 0.653 | 0.536 | 0.713 | 0.680 | 0.622 | 0.639 | 0.570 | 0.496 | 0.582 | 0.528 | 0.461 |
| PhyCMAP(C$_\beta$) | 0.741 | 0.641 | 0.534 | 0.746 | 0.653 | 0.611 | 0.655 | 0.571 | 0.500 | 0.636 | 0.550 | 0.477 |
| PSICOV(C$_\alpha$) | 0.459 | 0.343 | 0.258 | 0.528 | 0.469 | 0.417 | 0.462 | 0.363 | 0.282 | 0.483 | 0.418 | 0.358 |
| PSICOV(C$_\beta$) | 0.582 | 0.422 | 0.314 | 0.733 | 0.650 | 0.569 | 0.647 | 0.496 | 0.371 | 0.674 | 0.584 | 0.495 |
| NNCON(C$_\alpha$) | 0.475 | 0.390 | 0.318 | 0.377 | 0.313 | 0.267 | 0.342 | 0.284 | 0.236 | 0.224 | 0.182 | 0.152 |
| CMAPpro(C$_\alpha$) | 0.643 | 0.519 | 0.412 | 0.689 | 0.618 | 0.554 | 0.580 | 0.511 | 0.439 | 0.527 | 0.469 | 0.416 |
| CMAPpro(C$_\beta$) | 0.642 | 0.520 | 0.422 | 0.653 | 0.580 | 0.515 | 0.573 | 0.494 | 0.421 | 0.504 | 0.444 | 0.396 |
| Evfold(C$_\alpha$) | 0.382 | 0.297 | 0.235 | 0.488 | 0.442 | 0.398 | 0.451 | 0.366 | 0.289 | 0.442 | 0.389 | 0.342 |
| Accuracy on the 84 Set600 proteins with Meff ≤100 and ≤90% sequence identity with Astral 1.73 | | | | | | | | | | | | |
| PhyCMAP(C$_\alpha$) | 0.580 | 0.488 | 0.404 | 0.481 | 0.430 | 0.357 | 0.476 | 0.417 | 0.335 | 0.204 | 0.179 | 0.159 |
| PhyCMAP(C$_\beta$) | 0.548 | 0.468 | 0.392 | 0.454 | 0.408 | 0.345 | 0.452 | 0.399 | 0.331 | 0.220 | 0.214 | 0.187 |
| PSICOV(C$_\alpha$) | 0.070 | 0.071 | 0.072 | 0.065 | 0.050 | 0.044 | 0.074 | 0.055 | 0.049 | 0.063 | 0.043 | 0.035 |
| PSICOV(C$_\beta$) | 0.081 | 0.078 | 0.083 | 0.088 | 0.068 | 0.059 | 0.092 | 0.066 | 0.059 | 0.076 | 0.058 | 0.046 |
| NNCON(C$_\alpha$) | 0.535 | 0.421 | 0.342 | 0.324 | 0.298 | 0.248 | 0.348 | 0.321 | 0.271 | 0.162 | 0.132 | 0.114 |
| CMAPpro(C$_\alpha$) | 0.465 | 0.370 | 0.316 | 0.346 | 0.328 | 0.285 | 0.360 | 0.332 | 0.286 | 0.173 | 0.169 | 0.158 |
| CMAPpro(C$_\beta$) | 0.447 | 0.367 | 0.321 | 0.346 | 0.320 | 0.287 | 0.366 | 0.331 | 0.290 | 0.191 | 0.189 | 0.176 |
| Evfold(C$_\alpha$) | 0.074 | 0.068 | 0.066 | 0.079 | 0.058 | 0.039 | 0.074 | 0.053 | 0.045 | 0.063 | 0.042 | 0.032 |

### 3.1 Contribution of contrastive mutual information and pairwise contact scores

The contrastive mutual information (CMI) and homologous pairwise contact scores (HPS) help improve the performance of our random forest model. Table 8 shows their contribution to the prediction accuracy on the 471 proteins (with $M_{eff}$>100) in Set600.

**Table 8.** The contribution of contrastive mutual information and homologous pair contact scores to C$_\beta$ contact prediction. Similar results are observed for C$_\alpha$ contact prediction.

| | Short-range contacts | | | Medium and long-range | | |
|---|---|---|---|---|---|---|
| | Top 5 | L/10 | L/5 | Top 5 | L/10 | L/5 |
| with CMI and HPS | 0.754 | 0.632 | 0.521 | 0.720 | 0.649 | 0.589 |
| no CMI and HPS | 0.600 | 0.570 | 0.487 | 0.538 | 0.560 | 0.506 |

### 3.2 Contribution of the powers of the MI matrix

The powers of the MI matrix greatly improve long-range contact prediction when many sequence homologs are available, as shown in Table 9. The average accuracy is computed on the Set600 proteins with $M_{eff}$>100 excluding those with >90% sequence identify with Astral1.73. Figure 9 shows gain from using the powers of the MI matrix for each test protein. This result indicates that it is important to con-sider the chaining effect of residue couplings for medium- and long-range contacts and the powers of the MI matrix are good measure of the chaining effect. It is also worth to mention that we can calculate the powers of the MI matrix quickly, much faster than calculating the matrix inverse (as done by PSICOV).

**Table 9.** The contribution of the powers of the MI matrix to $C_\beta$ contact prediction, tested on the Set600 proteins with $M_{eff} > 100$.

|  | Short-range | | Medium and Long range | | Medium-range | | Long-range | |
|---|---|---|---|---|---|---|---|---|
|  | L/10 | L/5 | L/10 | L/5 | L/10 | L/5 | L/10 | L/5 |
| w/o the MI powers | 0.548 | 0.484 | 0.589 | 0.533 | 0.510 | 0.443 | 0.459 | 0.395 |
| with the MI powers | 0.556 | 0.484 | 0.651 | 0.592 | 0.564 | 0.491 | 0.550 | 0.477 |

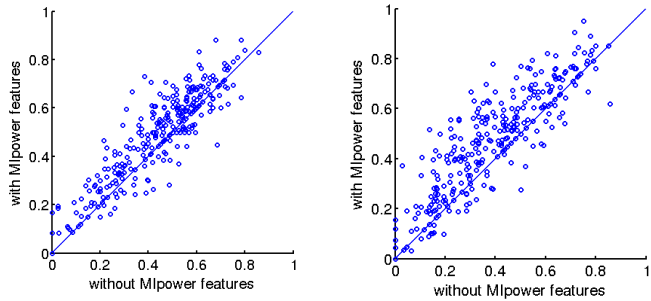

Figure 9. This figure shows the gain from the powers of the MI matrix. The first and second panels show the medium-range and long-range contact prediction, respectively. Each point represents the accuracy of the top L/5 predicted contacts by two Random Forest models, one with the powers of the MI matrix and the other without the MI powers.

### 3.3 Contribution of physical constraints

Table 10 shows the improvement resulting from the physical constraints (i.e., the ILP method) over the Random Forests (RF) method on Set600. On the 471 proteins with $M_{eff}$>100, ILP improves medium- and long-range contact prediction, but not short-range contact prediction. This result confirms that the physical constraints used by our ILP method indeed capture some global properties of a protein contact map. The improvement resulting from the physical constraints is larger on the 130 proteins with $M_{eff} \leq$100. In particular, the improvement on short-range contacts is substantial. These results may imply that when homologous information is sufficient, we can predict short-range contacts very accurately and thus, cannot further improve the prediction by the physical constraints. When homologous information is insufficient for accurate contact prediction, we can improve the prediction using the physical constraints, which are complementary to evolutionary information.

**Table 10.** The contribution of physical constraints for $C_\beta$ contact prediction.

|  | Short-range contacts | | | Medium and long-range | | |
|---|---|---|---|---|---|---|
|  | Top 5 | L/10 | L/5 | Top 5 | L/10 | L/5 |
| 471 Set600 proteins with $M_{eff} > 100$ | | | | | | |
| RF+ILP | 0.746 | 0.637 | 0.531 | 0.731 | 0.656 | 0.608 |
| RF | 0.754 | 0.632 | 0.521 | 0.720 | 0.649 | 0.589 |
| 130 Set600 proteins with $M_{eff} \leq 100$ | | | | | | |
| RF+ILP | 0.505 | 0.435 | 0.365 | 0.418 | 0.364 | 0.314 |
| RF | 0.445 | 0.368 | 0.299 | 0.414 | 0.342 | 0.281 |

### 3.4 Specific Examples

We show the contact prediction results of four CASP10 targets T0677D2, T0693D2, T0756D1 and T0701D1 in Figure 10, Figure 101, Figure 12, and Figure 13, respectively. T0677D2 does not have many sequence homologs ($M_{eff}$=31.53). As shown in Figure 10, our prediction matches well with the native contacts. PhyCMAP has top L/5 prediction accuracy 0.357 on medium- and long-range contacts while Evfold cannot correctly predict any contacts. T0693D2 has many sequence homologs with $M_{eff}$=2208.39. As shown in Figure 11, Phy-CMAP does well in predicting the long-distance contacts around the residue pair (20,100). For this target, PhyCMAP and Evfold obtain top L/5 prediction accuracy of 0.628 and 0.419, respectively, when both medium- and long-range contacts are considered. In Figure 12, T0756D1 has many sequence homologs with $M_{eff}$ = 1824.46. The accuracy of the top L/5 predicted contacts by our PhyCMAP is 0.944, while the accuracy by Evfold is 0.500. In Figure 13, T0701D1 has a large number of sequence homologs with $M_{eff}$ = 3300.02, the accuracy of the top L/5 predicted contacts by our PhyCMAP is 0.794, while the accuracy by Evfold is 0.444.

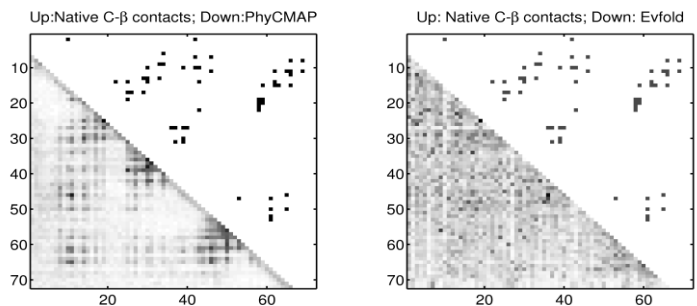

Figure 10. The predicted medium and long-range contacts for T0677D2. The upper triangles display the native $C_\beta$ contacts. The lower triangles of the left and right plots display the contact probabilities predicted by PhyCMAP and Evfold, respectively.

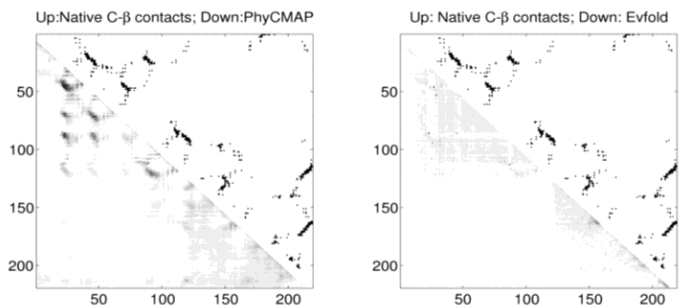

Figure 11. The predicted medium and long-range contacts for T0693D2. The upper triangles display the native $C_\beta$ contacts. The lower triangles of the left and right plots display the contact probabilities predicted by PhyCMAP and Evfold, respectively.

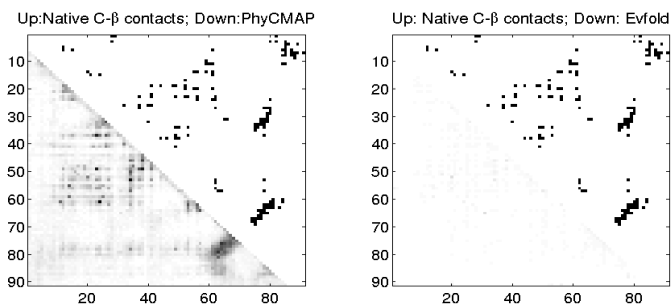

Figure 12. The predicted medium and long-range contacts for T0756D1. The upper triangles display the native $C_\beta$ contacts. The lower triangles of the left and right plots display the contact probabilities predicted by PhyCMAP and Evfold, respectively.

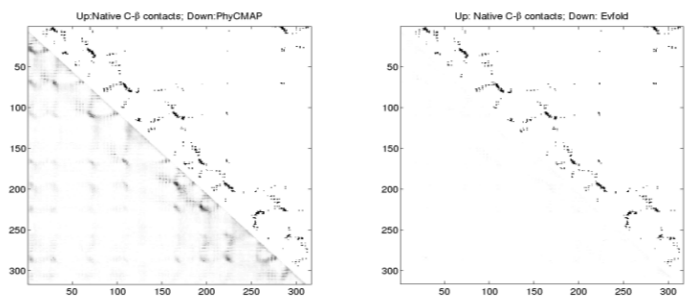

Figure 13. The predicted medium and long-range contacts for T0701D1. The upper triangles display the native C$_\beta$ contacts. The lower triangles of the left and right plots display the contact probabilities predicted by PhyCMAP and Evfold, respectively.

## 4 CONCLUSION AND DISCUSSIONS

This paper has presented a novel method for protein contact map prediction by integrating evolutionary information, non-evolutionary information and physical constraints using machine learning and integer linear programming. Our method differs from currently popular contact prediction methods in that we enforce a few physical constraints, which imply the sparsity constraint (used by PSICOV and Evfold), to the whole contact map and take into consideration contact correlation. Our method also differs from the first-principle method (e.g., Astro-Fold) in that we exploit evolutionary information from several aspects (e.g., two types of mutual information, context-specific distance potential and sequence profile) to significantly improve prediction accuracy. Experimental results confirm that our method outperforms existing popular machine learning methods (e.g., CMAPpro and NNCON) and two recent residue co-evolution methods PSICOV and Evfold regardless of the number of sequence homologs available for the protein under consideration.

The study of our method indicates that the physical constraints are helpful for contact prediction, especially when the protein under consideration does not have many sequence homologs. Nevertheless, the physical constraints used by our current method do not cover all the properties of a protein contact map. To further improve prediction accuracy on medium and long-range contact prediction, we may take into consideration global properties of a protein distance matrix. For example, the pairwise distances of any three residues shall satisfy the triangle inequality. Some residues also have correlated pairwise distances. To enforce this kind of distance constraints, we shall introduce distance variables and also define their relationship with contact variables. By introducing distance variables, we may also optimize the distance probability, as opposed to the contact probability used by our current ILP method. Further, we can also introduce variables of beta-sheet (group of beta-strands) to capture more global properties of a contact map.

One may ask how our approach compares to a model-based filtering strategy in which 3D models are built based on initial predicted contacts and then used to filter incorrect predictions. Our method differs from this general "model-based filtering" strategy in a couple of aspects. First, it is time-consuming to build thousands or at least hundreds of 3D models with initial predicted contacts. In contrast, our method can do contact prediction (using physical constraints) within minutes after PSI-BLAST search is done. Second, the quality of the 3D models is also determined by other factors such as energy function and energy optimization (or conformation sampling) methods while our method is independent of these factors. Even if the energy function is very accurate, the energy optimization algorithm often is trapped to local minima because the energy function is not rugged. That is, the 3D models resulting from energy minimization are biased towards a specific region of the conformation space, unless an extremely large scale of conformation sampling is conducted. Therefore, the predicted contacts derived from these models may also suffer from this "local minima" issue. By contrast, our integer programming method can search through the whole conformation space and find the global optimal solution and thus, is not biased to any local minima region. By using our predicted contacts as constraints, we may pinpoint to the good region of a conformation space (without being biased by local minima), reduce the search space and significantly speed up conformation search.

## 5 ACKNOWLEDGEMENTS

The authors are grateful to anonymous reviewers for their insightful comments and to the University of Chicago Beagle team and TeraGrid for their support of computational resources.

*Funding*: This work is supported by the National Institutes of Health grant R01GM0897532, National Science Foundation DBI-0960390, the NSF CAREER award and the Alfred P. Sloan Research Fellowship.